\documentclass[prd,preprint,tightenlines,floatfix,preprintnumbers,showpacs,nofootinbib]{revtex4}
 \usepackage[dvips,final]{graphicx}
  \usepackage{amssymb}
   \usepackage{amsmath}
    \usepackage{epsfig}
     \usepackage{bm}
      \usepackage{pifont}

\begin{document}
\thispagestyle{empty}
\date{\today}
\preprint{\vbox{\hbox{UA/NPPS-01-01}\hbox{RUB-TPII-22/00}}}
\title{Analyticity and power corrections in hard-scattering hadronic
       functions\\}
\author{A. I. Karanikas}
 \email{akaranik@cc.uoa.gr}
  \affiliation{University of Athens, Department of Physics,
                  Nuclear and Particle Physics Section,
                  Panepistimiopolis, GR-15771 Athens, Greece\\}
 \author{N. G. Stefanis}
  \email{stefanis@tp2.ruhr-uni-bochum.de}
   \affiliation{Institut f\"ur Theoretische Physik II,
                Ruhr-Universit\"at Bochum, D-44780 Bochum, Germany\\}

\begin{abstract}
Demanding the analyticity of hadronic observables (calculated in
terms of power series of the running coupling) as a {\it whole},
we show that they are free of the Landau singularity. Employing
resummation and dispersion-relation techniques, we compute in a
unifying way power corrections to two different hard-scattering
functions in perturbative QCD: the electromagnetic pion form factor
to leading order and the inclusive cross section of the Drell-Yan
process. In the second case, the leading nonperturbative power
correction in $b\Lambda_{\rm QCD}$ gives rise to a Sudakov-like
exponential factor in the impact parameter space which provides
enhancement rather than suppression.
\end{abstract}
\maketitle
\newpage   
\noindent
{\bf 1. Introductory remarks} \\
\indent
One of the most crucial questions in comparing QCD with experimental
data is whether predictions derived at the parton level can claim
validity also at the hadron level.
Would we be able to calculate binding (confinement) effects of quarks
and gluons reliably, this problem would be only marginal because then
calculated (i.e., partonic) and physical observables would almost
coincide.
Lacking such a theoretical scheme, the strategy must be to make
perturbative predictions less sensitive to nonperturbative effects.
Since parton binding effects become important at large distances (i.e.,
small momenta), we must ensure that sensitivity of hadronic observables
to the large-distance domain is minimized (infrared safety).
For such infrared-improved observables, one may expect that deviations
between theory and experiment, originating from the differences between
partons and hadrons, become practically irrelevant.

Normally, large-distance effects are factorized into universal,
wave-function parts that cannot be treated perturbatively.
However, infrared (IR) sensitivity may also reside in the initially
separated hard part that should describe partonic subprocesses
involving by definition hard propagators and hard vertices.
In fact, beyond leading order, this assumption cannot a priori be
satisfied because the momentum flow in such Feynman diagrams may become
much smaller than the large scale of the process, say, the
momentum transfer $Q^{2}$.
More precisely, the average gluon momentum $<k>$ flowing inside a
partonic subprocess decreases with the order of the perturbation
expansion and may eventually become proportional to
$\Lambda _{\rm QCD}$, implying that the resummed perturbation series
is defined only up to a certain power accuracy.
To make the result of the perturbative calculation unambiguous, one has
to compensate this power correction in the perturbative sector by a
corresponding term of the same form, originating from the
nonperturbative regime.

In processes which involve the emission of virtual gluon quanta of low
momentum, the strategy must be to resum their contributions to all
orders of the strong coupling constant.
This gives rise to exponentially suppressing factors of the reaction
amplitude (or cross-section) of the Sudakov type with exponents
containing double and single logarithms of the large mass scale.
However, because of the Landau singularity of the running coupling at
transverse distances $b\propto 1/\Lambda _{\rm QCD}$ (where $b$ is the
impact parameter conjugate to the parton's transverse momentum
$k_{\perp}$), an essential singularity appears in the Sudakov factor.
Thus, one has to consider power corrections of
${\cal O}\left( b^{2}\Lambda _{\rm QCD}^{2} \right)$, which, though
negligible for small $b$ relative to logarithmic corrections
$\ln\left( b^{2}\Lambda _{\rm QCD}^{2} \right)$,
may become important for larger values of the impact parameter.

In this letter we will describe a general methodology to treat
power series in the running strong coupling. To be more precise,
we will address these questions having recourse to two processes:
one to which the OPE applies, namely the pion electromagnetic form
factor at leading perturbative order, and another, the Drell-Yan
process, to which the OPE is not applicable. The first one is a
typical example of an exclusive process with registered hadrons in
the initial and final states (for a recent review, see, e.g.,
\cite{Ste99}). The Drell-Yan mechanism, on the other hand, has two
identified hadrons in the initial state and a lepton pair (plus
unspecified particles) in the final state, whose transverse
momentum distribution is proportional to the large invariant mass
of the materialized photon.

Our goal in the second case is to obtain not only the usual resummed
(Sudakov) expression, comprising logarithmic corrections due to
soft-gluon radiation, but also to include the leading power correction
as well, specifying, in particular, its concomitant coefficient.
This becomes possible within a theoretical scheme, which models the IR
behavior of the running coupling by demanding analyticity of physical
observables (in the complex $Q^{2}$ plane) as a {\it whole} -- as
opposed to imposing analyticity of individual powers, i.e., order by
order in perturbation theory --, while preserving renormalization-group
invariance (references and additional information can be found in the
recent surveys \cite{Shi00,SS99}).

Our own physical viewpoint can be summarized as follows:
One option to access nonperturbative effects is to change degrees of
freedom and replace QCD by some low-energy effective theory.
The other option, and that actually adopted here, is to retain the
usual QCD degrees of freedom, but to demand that hadronic observables,
calculated perturbatively with them, are analytic as a whole in the
$Q^{2}$ plane.
This analytization condition entails a singularity-free expression for
the strong running coupling in both the spacelike and the timelike
region, rendering hadronic observables IR safe.
In this way, we are able to calculate explicitly in our second example
the $b^{2}\Lambda _{\rm QCD}^{2}$ power correction to the Drell-Yan
cross-section and show that after exponentiation it amounts to a
Sudakov type factor that can compete in magnitude with the resummed
double logarithms due to gluonic radiative corrections because it has
the reverse sign, whence providing enhancement rather than suppression.
The properties of this new factor are addressed and its (almost)
Gaussian dependence on $b$ is discussed and compared to previous works
\cite{KS95,ASS98}.

\bigskip

\noindent
{\bf 2. Power corrections to pion form factor} \\
\indent
Following the rationale of analyticity in the sense just described, the
leading-order factorized pion form factor reads \cite{ER80,LB80}
\begin{equation}
  \left[ Q^{2}F_{\pi} \left( Q^{2} \right) \right]_{\rm an}
=
  \int_{0}^{1} dx
  \int_{0}^{1} dy
  \left[ \Phi _{\pi}^{\rm out} \left( y, Q^{2} \right)
  T_{\rm H}
  \left(
        x, y, Q^{2}, \alpha _{s} \left( \hat{Q}^{2} \right)
  \right)
  \Phi _{\pi}^{\rm in} \left( x, Q^{2} \right)
  \right]_{\rm an} \; ,
\label{eq:ffanaly}
\end{equation}
where $\Phi _{\pi}\left( x, Q^{2} \right)$ is the process-independent
pion distribution amplitude, encoding the nonperturbative binding
dynamics of the valence quarks at the resolution scale $Q^{2}$, each
carrying light-cone momentum fractions
$x_{1}  = x$ (quark) and
$x_{2}  = 1 - x\equiv \bar x$ (antiquark)
$(x_{i} = k_{i}^{+}/P^{+})$
of the pion's momentum $P^{\mu}$, and the superscripts in and out
denote, respectively, incident (incoming) and final (outgoing) intact
pions.

Employing the asymptotic pion distribution amplitude, evolution
effects can be neglected \cite{Mue94} and the analyticity
requirement resides only in the hard-scattering part. Hence, we
have
\begin{equation}
  \left[ Q^{2}F_{\pi}\left( Q^{2} \right) \right]_{\rm an}
=
  A
  \int_{0}^{1} dx
  \int_{0}^{1} dy
  xy \bar{x}\bar{y}
  \left[
        T_{\rm H}
        \left(
              x, y, Q^{2}, \alpha _{s}\left( \hat{Q}^{2} \right)
        \right)
        \right]_{\rm an} \; ,
\label{eq:analyhard}
\end{equation}
where to leading perturbative order, the hard-scattering
amplitude is given by
\begin{equation}
  T_{\rm H}^{(1)}
  \left(
        x, y, Q^{2}, \alpha _{s}\left( \hat{Q}^{2} \right)
  \right)
=
  16 \pi C_{F}
  \left[
        \frac{2}{3}
  \frac{\alpha _{s}\left( Q^{2}\bar{x}\bar{y} \right)}{\bar{x}\bar{y}}
+
  \frac{1}{3}
        \frac{\alpha _{s}\left( Q^{2}xy \right)}{xy}
  \right]
\label{eq:hardpart}
\end{equation}
and the constant $A$ takes account of the correct normalization of the
pion distribution amplitude.
Had we not imposed the requirement of analytization, expression
(\ref{eq:analyhard}) would lead to an asymptotic series in the coupling
constant
$\alpha _{s}\left( Q^{2} \right)$
that is not Borel summable, as noticed by Agaev \cite{Aga96} (see also
\cite{Aga98}).

Global analytization \cite{Shi00,Shi99} (see also \cite{GI99,BRS00})
now means that
\begin{equation}
  {\left[
         \alpha _{s}^{n}\left( Q^{2} \right)
  \right]}_{\rm an}
\equiv
  \frac{1}{\pi}
  \int_{0}^{\infty}
  \frac{d\xi}{\xi + Q^{2} - i \epsilon}
  \rho ^{(n)}(\xi ) \; ,
\label{eq:analycond}
\end{equation}
where the spectral density $\rho ^{(n)}(\xi )$ is the dispersive
conjugate of all powers $n$ of $\alpha _{s}$.
For the leading-order expression under consideration the spectral
density is \cite{Rad82,KP82,Piv92,SS97}
\begin{equation}
  \rho \left( Q^{2} \right)
=
  {\rm Im} \alpha _{s}\left( - Q^{2} \right)
=
  \frac{\pi}{\beta _{1}}
  \frac{1}{\ln ^{2}\frac{Q^{2}}{\Lambda ^{2}} + \pi ^{2}}
\label{eq:specdens}
\end{equation}
with $\Lambda \equiv \Lambda _{\rm QCD}$.
Then Eq.~(\ref{eq:analycond}) reduces to
\begin{equation}
  \left[ \alpha _{s}\left( Q^{2} \right) \right]_{\rm an}
=
  \frac{1}{2\pi i}
  \int_{C}^{} \frac{dz}{z - Q^{2} + i \epsilon}
  \alpha _{s}(z) \; ,
\label{eq:alphanal}
\end{equation}
where $C$ is a closed contour in the complex $z$-plane with a
branch cut along the negative real axis, so that
($\beta _{1} = \beta _{0}/4\pi$ with $\beta _{0} = 11 - 2 N_{f}/3$)
\begin{equation}
  {\left[
         \alpha _{s}\left( Q^{2} \right)
  \right]}_{\rm an}
=
  \frac{1}{\beta _{1}}
  \left(
          \frac{1}{\ln \frac{Q^{2}}{\Lambda ^{2}}}
        + \frac{1}{1 - \frac{Q^{2}}{\Lambda ^{2}}}
  \right) \; ,
\label{eq:alphasan}
\end{equation}
an expression recently proposed by Shirkov and Solovtsov \cite{SS97}.

Recasting now the strong coupling in the form
\begin{equation}
  \alpha _{s}(z)
=
  \frac{1}{\beta _{1}}
  \frac{1}{\ln \frac{z}{\Lambda ^{2}}}
=
  \pm \int_{0}^{\infty} d\sigma
      \exp{\left( \mp\sigma \beta _{1} \ln \Lambda ^{2}/z \right)} \; ,
\label{eq:expalpha}
\end{equation}
with the plus sign corresponding to the case $|z|/\Lambda ^{2}>1$ and
the minus one to $|z|/\Lambda ^{2}<1$, and inserting it into
Eq.~(\ref{eq:analyhard}), we find after some standard manipulations
the Borel transform of the scaled pion form
factor at leading perturbative order:
\begin{equation}
  \left[ Q^{2}F_{\pi}\left( Q^{2} \right) \right]_{\rm an}^{(1)}
=
  \int_{0}^{\infty}
  d\sigma
  \exp{\left( - \sigma \beta _{1} \ln Q^{2}/\Lambda ^{2} \right)}
  \tilde{\pi}(\sigma )_{\rm an}^{(1)} \; .
\label{eq:borelff}
\end{equation}
Here the Borel image of the form factor reads
\begin{eqnarray}
  \tilde{\pi}(\sigma )_{\rm an}^{(1)}
= &&
  16 \pi C_{F} A
  \frac{\sin \left(\pi \beta _{1}\sigma \right)}{\pi}
  \int_{0}^{1} dx
  \int_{0}^{1} dy
  \bar{x}\bar{y}
  \int_{0}^{\infty}
  \frac{d\xi}{\xi + xy}
\nonumber \\
&&
  \Biggl[
  \xi ^{-\sigma\beta _{1}}
     \Theta \left( \xi - \frac{\Lambda ^{2}}{Q^{2}} \right)
   + \left( \frac{Q^{2}}{\Lambda ^{2}} \right)^{2\sigma\beta _{1}}
     \xi ^{\sigma\beta _{1}}
     \Theta \left( \frac{\Lambda ^{2}}{Q^{2}} - \xi \right)
  \Biggr] \; .
\label{eq:boreltrans}
\end{eqnarray}
We stress that this expression has no IR renormalons in contrast
to approaches which use the conventional one-loop $\alpha _{s}$
parameterization (see, e.g., \cite{Aga96,Aga98}). Carrying out the
integrations, we then obtain the following final result
\begin{eqnarray}
  \tilde{\pi}(\sigma )_{\rm an}^{(1)}
& = &
  16 \pi C_{F} A
  \frac{\sin \left( \pi \beta _{1}\sigma \right)}{\pi \beta _{1}\sigma}
  \left( \frac{\Lambda ^{2}}{Q^{2}} \right)^{1 - \beta _{1}\sigma}
  \int_{0}^{Q^{2}/\Lambda ^{2}}
  \frac{dw}{1+w}
  \phi \left( \frac{w\Lambda ^{2}}{Q^{2}} \right)\nonumber\\
&&\times  \Biggl[
         {}_{2}F_{1}\left( 1,1;1+\beta _{1}\sigma;\frac{w}{1+w} \right)
      +  \frac{\beta _{1}\sigma}{1+\beta _{1}\sigma}
         {}_{2}F_{1}\left( 1,1;2+\beta _{1}\sigma;\frac{1}{1+w} \right)
  \Biggr]
\label{eq:finpiborel}
\end{eqnarray}
with ${}_{2}F_{1}$ being the hypergeometric function and $\phi (w)$
denoting the abbreviation
\begin{equation}
    \phi (w)
=
  - (1+w)\, \ln w - 2(1-w)
\quad\quad\quad\quad
    (\phi (w)\geq 0 \quad \mbox{\rm when} \quad w\leq 1) \; .
\end{equation}
\label{eq:fiofw}

Inserting this expression into Eq.~(\ref{eq:borelff}), the integration
over the Borel parameter $\sigma$ can be performed without any
ambiguity to arrive at the following result for the pion form factor
(in leading order)
\begin{equation}
  {\left[ Q^{2}F_{\pi}\left( Q^{2} \right) \right]}_{\rm an}^{(1)}
=
  16 \pi C_{F} A
  \frac{1}{\beta _{1}}
  \int_{0}^{1} dw\, \phi (w)
  \left[
         \frac{1}{\ln \left( \frac{wQ^{2}}{\Lambda ^{2}} \right)}
        +
         \frac{1}{1-\frac{wQ^{2}}{\Lambda ^{2}}}
  \right] \; .
\label{eq:pifofa1}
\end{equation}
Hence, the remaining integration can be carried out analytically and
the final result is
\begin{eqnarray}
  {\left[ Q^{2}F_{\pi}\left( Q^{2} \right) \right]}_{\rm an}^{(1)}
& = &
  16 \pi C_{F} A
  \frac{1}{\beta _{1}}
 \Biggl\{
          - \frac{3}{2}
          + \frac{\Lambda ^{2}}{Q^{2}}
          \left[
                \ln \left( \frac{Q^{2}}{\Lambda ^{2}} \right) - 2
          \right]
          \left[
                  {\rm Li}_{2}\left( \frac{Q^{2}}{\Lambda ^{2}} \right)
                - \ln \left( \frac{Q^{2}}{\Lambda ^{2}} - 1 \right)
          \right]
\nonumber \\
&& +
          \frac{\Lambda ^{4}}{Q^{4}}
          \left[
                 \ln \left( \frac{Q^{2}}{\Lambda ^{2}}\right) + 2
          \right]
          \left[
                  {\rm Li}_{2}\left( \frac{Q^{4}}{\Lambda ^{4}} \right)
                - \ln \left( \frac{Q^{2}}{\Lambda ^{2}} - 1 \right)
          \right]
\nonumber \\
&& +
          \frac{\Lambda ^{2}}{Q^{2}}
          \left(
                1 + \frac{\Lambda ^{2}}{Q^{2}}
          \right)
          \left[
                  \frac{1}{2}
                  \ln ^{2} \left( \frac{Q^{2}}{\Lambda ^{2}} \right)
                + \frac{\pi ^{2}}{6} - 1
          \right]
\nonumber \\
&& -
    2 \frac{\Lambda ^{2}}{Q^{2}} + \frac{\Lambda ^{4}}{Q^{4}}
    - \frac{\Lambda ^{2}}{Q^{2}}
      \left(
            1 + \frac{\Lambda ^{2}}{Q^{2}}
      \right)
      \sum_{n=1}^{\infty}
      \Biggl\{
              \frac{1}{n}
                 {\left( \frac{\Lambda ^{2}}{Q^{2}} \right)}^{n}
              \ln \left( \frac{Q^{2}}{\Lambda ^{2}} \right)
\nonumber \\
&& -
              \frac{1}{n^{2}}
              \left[
                    1 - {\left( \frac{\Lambda ^{2}}{Q^{2}} \right)}^{n}
              \right]
      \Biggr\}
 \Biggr\} \; .
\label{eq:pifofafin}
\end{eqnarray}
It is important to notice that the above expression remains
analytic all the way down to the limit $\frac{Q^{2}}{\Lambda
^{2}}\to 1$. This is to be contrasted with the corresponding
result found by Agaev in \cite{Aga96} (his equation (16)) which
comes out divergent in this limit and has to be regularized.

For the physically relevant case $Q^{2} \gg \Lambda ^{2}$,
Eq.~(\ref{eq:pifofafin}) simplifies to
\begin{eqnarray}
  {\left[ Q^{2}F_{\pi}\left( Q^{2} \right) \right]}_{\rm an}^{(1)}
& = &
  16 \pi C_{F} A
  \left[
          \frac{1}{4} \alpha _{s}\left( Q^{2} \right)
        + {\cal O}\left( \alpha _{s}^{2} \right)
  \right]
\nonumber \\
&& -
   \frac{1}{\beta _{1}}
   16 \pi C_{F} A
   \frac{\Lambda ^{2}}{Q^{2}}
   \left[
           \frac{1}{2}\ln ^{2}\left( \frac{Q^{2}}{\Lambda ^{2}} \right)
         - 2          \ln     \left( \frac{Q^{2}}{\Lambda ^{2}} - 1 \right)
         - \frac{\pi ^{2}}{3} + 3
   \right]\nonumber \\
&&  + {\cal O}\left( \frac{\Lambda ^{4}}{Q^{4}} \right) \; ,
\label{eq:pifofaQlarge}
\end{eqnarray}
where ${\rm Li}_{2}$ is the di-logarithm (or Spence) function, defined
by
\begin{equation}
  {\rm Li}_{2}(x)
=
  \int_{0}^{x} \, \frac{dt}{\ln t} \; .
\label{eq:spence}
\end{equation}
The computation carried out above can be extended to any desired order
of the hard-scattering amplitude $T_{\rm H}$.
The resulting expressions are always well-defined without the need of
employing any (additional) IR regularization \cite{KSS01}.

\bigskip

\noindent
{\bf 3. Power corrections to Drell-Yan process} \\
\indent
As a second example of our methodology, we discuss the derivation of
power corrections to the inclusive Drell-Yan cross-section with the
large scale $Q^{2}$ being here the invariant lepton pair mass.
Power corrections to this process with the help of renormalons have
been discussed in \cite{KS95,QS91,CS93,AZ95,BB95}.
Let us start our discussion here by evaluating the logarithmic
derivative\footnote{We take the logarithmic derivative because
perturbative QCD predicts not the absolute magnitude of reaction
amplitudes, but only their variation with momentum.
The derivative also eliminates the collinear divergence related to the
integrations over $k^{\pm}$.} of the unrenormalized expression for the
eikonalized Drell-Yan cross-section \cite{KM93} (see also \cite{Ste87})
after exponentiation of infrared divergences to the lowest order of
perturbation theory, adopting here and below for the ease of comparison
the notations of Korchemsky and Sterman \cite{KS95}:
\begin{eqnarray}
  \frac{d\ln W_{\rm DY}}{d\ln Q^{2}}
& \equiv &
  \Pi ^{(1)}\left( Q^{2} \right)
\nonumber \\
& = &
  4 C_{F} \mu ^{2\epsilon}
  \int_{}^{} \frac{d^{2-2\epsilon}k_{\perp}}{(2\pi )^{2-2\epsilon}}
  \frac{1}{k_{\perp}^{2}} \,
  \alpha _{s}\left( k_{\perp}^{2} \right)
  \left(
        {\rm e}^{-i {\bf k}_{\perp} \cdot {\bf b}} - 1
  \right) \; .
\label{eq:DYcrossec}
\end{eqnarray}
The argument of the strong coupling is taken to depend on the
transverse momentum $k_{\perp}^{2}$ in order to account for
higher-order quantum corrections, originating from momentum scales
larger than this \cite{CG80,KT82}.
It is obvious that the above integral is not well-defined at very
small mass scales owing to the singularities of the one (or higher)
loop QCD running coupling in this region.
This makes its evaluation at the edge of phase space sensitive to their
regularization.
The effect of regularizing the $k_{\perp}$ integration in the infrared
region amounts to including power corrections to the original
perturbative result which are the footprints of soft gluon emission at
the kinematic boundaries to the non-perturbative QCD regime
\cite{SSK00}.

By the same reasoning as applied in the previously considered case,
we impose analytization as a whole, and using Eq.~(\ref{eq:expalpha}),
we perform the $k_{\perp}$ integration in (\ref{eq:DYcrossec}) to
obtain
\begin{equation}
  {\left[
         \Pi ^{(1)}\left( Q^{2} \right)
  \right]}_{\rm an}
=
  \int_{0}^{\infty} d\sigma \,
  {\rm e}^{- \sigma \beta _{1}
             \ln \left( 4/b^{2}\Lambda ^{2} \right)}
  \tilde{\Pi}_{\rm an}^{(1)} (\sigma )
\label{eq:anDY}
\end{equation}
with the Borel transform given by
\begin{eqnarray}
  \tilde{\Pi}_{\rm an}^{(1)}( \sigma )
&& =
  \frac{4 C_{F}}{\pi}
  \left(
        \frac{\mu ^{2}b^{2}}{4}
  \right)^{\epsilon}
  \sin \left(
             \pi \sigma \beta _{1}
       \right)
  \int_{0}^{\infty} d\xi \, g( \xi )
  \Biggl[
         \xi ^{- \sigma \beta _{1}}
         \Theta \left(
                       \xi - \frac{b^{2}\Lambda ^{2}}{4}
                \right)
\nonumber \\
&& +
         \left(
               \frac{b^{2}\Lambda ^{2}}{4}
         \right)^{- 2 \sigma \beta _{1}}
         \xi ^{ \sigma \beta _{1}}
         \Theta \left(
                      \frac{b^{2}\Lambda ^{2}}{4} - \xi
                \right)
  \Biggr] \; ,
\label{eq:anDYborel}
\end{eqnarray}
where
\begin{equation}
  g( \xi )
=
  \int_{}^{} \frac{d^{2-2\epsilon}q}{( 2\pi )^{2-2\epsilon}}
  \frac{1}{q^{2}}
  \frac{1}{q^{2} + \xi}
  \left(
        {\rm e}^{- 2 i {\bf q} \cdot {\hat{\bf b}}} - 1
  \right) \; .
\label{eq:propagator}
\end{equation}
Combining denominators in Eq.~(\ref{eq:propagator}) and carrying out
the integrations over $\xi$, we then find
\begin{eqnarray}
  {\left[
         \Pi ^{(1)} \left( Q^{2} \right)
  \right]}_{\rm an}
&& =
  \frac{C_{F}}{\pi}
  \left( \mu ^{2}b^{2}\pi
  \right)^{\epsilon}
  \!\int_{0}^{\infty} d\sigma \!
  {\rm e}^{- \sigma \beta _{1}
  \ln \left(
            4/b^{2}\Lambda ^{2}
      \right)}
  \frac{1}{\Gamma \left( 1 + \sigma \beta _{1} \right)}
  \Biggl[
         - \frac{1}{\sigma \beta _{1} + \epsilon}
           \Gamma \left( 1 - \sigma \beta _{1} - \epsilon
                  \right)
\nonumber \\
&& +     \sum_{n=0}^{\infty}
         \frac{(-1)^{n}}{(n + 1)!}
         \left(
               \frac{b^{2}\Lambda ^{2}}{4}
         \right)^{n + 1 - \sigma \beta _{1} - \epsilon}
         \frac{1}{n + 1 - \sigma \beta _{1} - \epsilon}
  \Biggr]
   -     \frac{C_{F}}{\pi \beta _{1}}
         f\! \left( \frac{b^{2}\Lambda ^{2}}{4}
           \right) \, ,
\label{eq:DYoversigma}
\end{eqnarray}
where
\begin{eqnarray}
  f\left( \frac{b^{2}\Lambda ^{2}}{4} \right)
&& =
  \sum_{n=0}^{\infty} \frac{1}{(n + 1)!}
  \left[
        \frac{(-1)^{n}}{2} + 1 + \frac{{\rm si}(\pi(n+1))}{\pi}
  \right]
  \Biggl[
         \left( \frac{b^{2}\Lambda ^{2}}{4} \right)^{n+1}
         \Gamma \left( -n -1, \frac{b^{2}\Lambda ^{2}}{4} \right)
\nonumber \\
&& +     \frac{1}{n+1}
         \left( \frac{b^{2}\Lambda ^{2}}{4} \right)^{n+1}
         \Gamma \left( -n -2, 1 \right) - \frac{1}{n+1}
  \Biggr]
\nonumber \\
&& + \sum_{n=1}^{\infty} \frac{1}{(n!)^{2}}
     \left( \frac{b^{2}\Lambda ^{2}}{4} \right)^{n}
     \left[
           - \frac{1}{2} \ln \frac{b^{2}\Lambda ^{2}}{4}
           + \psi (n+1)
     \right]
\nonumber \\
&& + \sum_{n=1}^{\infty} \frac{1}{(n!)^{2}}
     (-1)^{n+1} \frac{2{\rm si}(\pi n)}{\pi}
     \left( \frac{b^{2}\Lambda ^{2}}{4} \right)^{n}
  \left[
        - \frac{1}{2} \ln \frac{b^{2}\Lambda ^{2}}{4}
        + \frac{1}{2} \frac{1}{n+1} + \psi (n+1)
  \right]
\label{eq:expf}
\end{eqnarray}
with $\Gamma (x,y)$, denoting the incomplete Gamma function
\cite{GR80},
\begin{equation}
  {\rm si}(x)
=
  - \int_{x}^{\infty} \frac{\sin t}{t} dt
=
  - \frac{\pi}{2} + \int_{0}^{x}\frac{\sin t}{t} dt \; ,
\label{eq:sinint}
\end{equation}
and
$
 \psi (x)
=
 (d/dx) \ln \Gamma (x)
$.

At this point some important remarks are in order.
The first term in Eq.~(\ref{eq:DYoversigma}), viz., the integral
over $\sigma$, diverges for $\sigma \beta _{1}=0$, i.e., for small
values of $\alpha _{s}\left( k_{\perp} \right)$ (or equivalently for
large transverse momenta $k_{\perp}$).
This ultraviolet (UV) divergence is taken care of by the defect of
the dimension $\epsilon$ within the $\overline{MS}$ renormalization
scheme we have adopted.
Were it not for the terms containing powers of $b \Lambda$, our
expression (\ref{eq:DYoversigma}) and that found by Korchemsky and
Sterman in \cite{KS95} (namely, their equation (18)) would be the same.
In our case, however, the imposition of analytization cures all
divergences related to IR renormalons that are generated by the
$\Gamma$-functions whenever $\sigma\beta _{1}$ is an integer different
from zero.
On the other hand, when $\sigma\beta _{1}$ is an integer different from
zero, say, $\sigma\beta _{1}=m$, then the integrand in
Eq.~(\ref{eq:DYoversigma}) takes the form
\begin{equation}
   \frac{(-1)^{m - 1}}{(m!)^{2}}
   \left( \frac{b^{2}\Lambda ^{2}}{4} \right)^{m}
   \ln \left(
             {\rm e}^{-B}\, \frac{b^{2}\Lambda ^{2}}{4}
       \right)
 + \sum_{n=0}^{\infty}\!\!{\phantom{\Bigl|}}^{\prime}
   \frac{(-1)^{n}}{(n + 1)!m!}
   \left(
   \frac{b^{2}\Lambda ^{2}}{4}
   \right)^{n+1}
   \frac{1}{n + 1 - m} \; ,
\label{eq:integer}
\end{equation}
where
\begin{equation}
  B
=
  1 - \gamma _{E} + \sum_{k=1}^{m-1} \frac{1}{k + 1}
\label{eq:constB}
\end{equation}
and the prime on the sum symbol reminds that the term $n = m - 1$
is excluded.
(Obviously, for $m=1$ the sums above vanish.)

Let us now turn our attention to the second term in
Eq.~(\ref{eq:DYoversigma}).
Evaluating further this term and retaining only the leading
contributions in $b^{2}\Lambda ^{2}$, we finally obtain
\begin{equation}
  f\left( b^{2}\Lambda ^{2} \right)
=
    - a_{0}
    - a_{1}\frac{b^{2}\Lambda ^{2}}{4}
           \ln\frac{b^{2}\Lambda ^{2}}{4}
    + a_{2}\frac{b^{2}\Lambda ^{2}}{4}
    + {\cal O}\left( b^{4}\Lambda ^{4} \right) \; ,
\label{eq:powersfin}
\end{equation}
in which the following abbreviations
\begin{eqnarray}
  a_{0}
& = &
  0
\nonumber \\
  a_{1}
& \equiv &
  A_{0} + \frac{1}{2} + \frac{{\rm si}(\pi )}{\pi}
\simeq 3.18
\nonumber \\
  a_{2}
& \equiv &
  - A_{0}\gamma_{\rm E} - \sum_{n=0}^{\infty}
    \left[ \frac{A_{n}}{n+1} + \frac{A_{n+1}}{(n+2)(n+1)}
    \right]
  + A_{0}\Gamma(-2,1)
  + \psi(2)
  + \frac{{2\rm si}(\pi )}{\pi}\left[\frac{1}{4} +
  \psi(2)\right]\nonumber\\
& \simeq &
  -2.51
\label{eq:coeff}
\end{eqnarray}
have been used with $\gamma _{E}=0.5772\ldots$ being Euler's
gamma function and
\begin{equation}
  A_{n}
=
  \frac{1}{(n+1)!}\left[\frac{(-1)^{n}}{2} + 1 + \frac{{\rm si}(\pi(n+1))}{\pi}
                  \right]\, .
\label{eq:A_n}
\end{equation}

Though the integral in Eq.~(\ref{eq:DYoversigma}) (i.e., the first
term) cannot be computed in closed form, it can be expanded in terms of
powers of $b^{2}\Lambda ^{2}$.
The only singularity of the integrand is a single UV pole at
$\sigma\beta _{1}=0$, which is regularized dimensionally.
For $\sigma\beta _{1}$ an integer, both terms inside the bracket have
poles, but they mutually cancel so that their sum is singularity free
and therefore the integral is well defined.
Retaining terms of order $b^{2}\Lambda ^{2}$, it is apparent from
expression (\ref{eq:integer}) that the main contribution stems from
the leading renormalon $\sigma\beta _{1}=1$.
The result is
\begin{equation}
  {\left[ \Pi ^{(1)}\left( Q^{2} \right) \right]}_{\rm an}
=
   {\left[ \Pi ^{(1)}\left( Q^{2} \right) \right]}_{\rm PT}
 + {\left[ \Pi ^{(1)}\left( Q^{2} \right) \right]}_{\rm pow}
\label{eq:DYfinal}
\end{equation}
with the perturbative part being defined by
\begin{equation}
   {\left[ \Pi ^{(1)}\left( Q^{2} \right) \right]}_{\rm PT}
=
   \frac{C_{F}}{\pi\beta _{1}}
   \ln \frac{\ln \left( C/b^{2}\Lambda ^{2} \right)}
            {\ln \left( Q^{2}/\Lambda ^{2} \right)} \; ,
\label{eq:DYfinpt}
\end{equation}
where, within the $\overline{MS}$ scheme, we have set for the
renormalization scale (the collinear limit)
$\mu ^{2} = 2Q^{2}$.
Notice that this leading-order result coincides with the one obtained
by Korchemsky and Sterman (Eq.~(20) in \cite{KS95}).
Power corrections in the impact parameter $b$ are encoded in the second
contribution ($b^{2}\Lambda ^{2}\ll 1$):
\begin{equation}
  {\left[ \Pi ^{(1)}\left( Q^{2} \right) \right]}_{\rm pow}
=
    S_{0} + b^{2} S_{2}\left( b^{2}\Lambda ^{2} \right)
  + {\cal O}\left( b^{4}\Lambda ^{4}
\right)
\;
,
\label{eq:DYpow}
\end{equation}
where
\begin{equation}
  S_{0}
=
  \frac{C_{F}}{\pi\beta _{1}} a_{0} = 0
\label{eq:powconst}
\end{equation}
and
\begin{equation}
  S_{2}\left( b^{2}\Lambda ^{2} \right)
 =
  \frac{C_{F}}{4\pi\beta _{1}} \Lambda ^{2}
    a_{1}\ln \frac{b^{2}\Lambda ^{2}}{4}
  - a_{2}
\label{eq:leadpow}
\end{equation}
with the constants $a_{0}$, $a_{1}$, and $a_{2}$ already introduced
in Eq.~(\ref{eq:coeff}).

\begin{figure}
\centerline{\epsfig{file=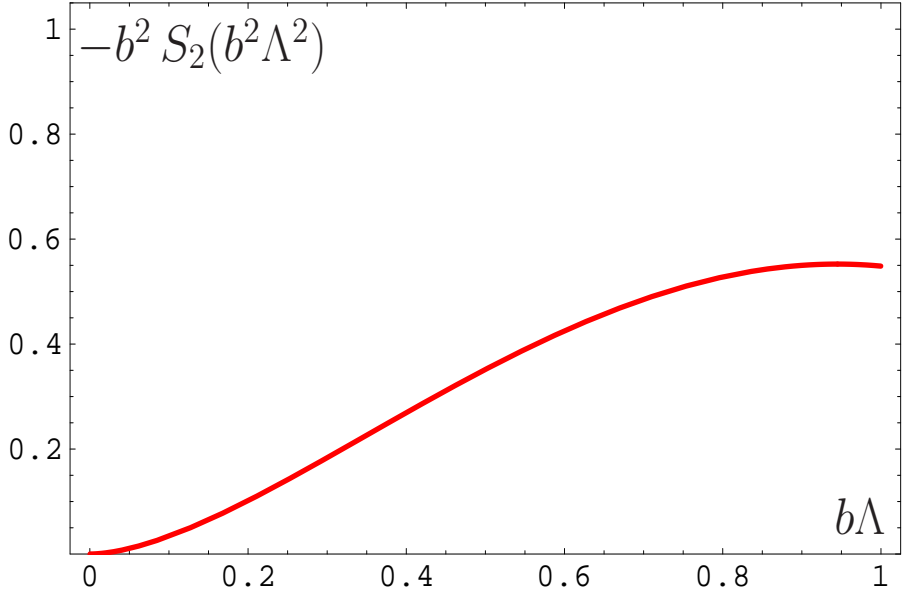,height=8.0cm,width=12.0cm,silent=}}
\vspace{0.5cm} \caption[fig:powers]
        {Exponent of the first non-constant power correction
        to the Drell-Yan cross-section as a function of the impact
        parameter $b$.
\label{fig:power2}}
\end{figure}

Hence, the Drell-Yan cross-section $W_{\rm DY}$, comprising the leading
logarithmic perturbative contribution (alias the leading Sudakov
exponent $S_{\rm PT}$) and including the first power correction in
$b^{2}\Lambda ^{2}$ reads
\begin{equation}
  W_{\rm DY}(b, Q)
=
  \exp \left[
             - S_{\rm PT}(b, Q) - b^{2} S_{2}(b,Q) + \ldots
       \right] \; ,
\label{eq:DYCSfinal}
\end{equation}
\begin{eqnarray}
  S_{2}(b,Q)
& \sim &
   S_{2}\left( b^{2}\Lambda ^{2} \right)\ln Q + \mbox{const} \, .
\label{eq:DYpowers}
\end{eqnarray}
Note that the $Q$-dependence arises due to collinear interactions,
i.e., through the integration of Eq.~(\ref{eq:DYcrossec}).
While the Sudakov factor, representing the perturbative tail of the
hadronic wave function \cite{Ste99,SSK00,Ste95}, suppresses constituent
configurations which involve large impact space separations, the
exponentiated power corrections in $b^{2}$ (leaving aside the constant
term $S_{0}$), which are of nonperturbative origin, provide enhancement
for such configurations, since
$S_{2}\left(b^{2}\Lambda ^{2}\right)$ (see Fig.~\ref{fig:power2})
is always negative.
The consequence is that combining (resummed) logarithmic radiative
corrections and power-behaved corrections in $b$, the latter arising
from soft (nonperturbative) gluon emission and being therefore
universal, the net result is less suppression of the Drell-Yan
cross-section \cite{KS95}.
Here we have an immediate link with the work of Akhoury et al.
\cite{ASS98} who pointed out within the renormalon approach \cite{KS95}
that the form of the exponent in Eq.~(\ref{eq:DYCSfinal}) is the same
as for the Fourier-transformed pion wave function.
In both cases the leading power correction in $b^{2}\Lambda ^{2}$ has
an exponential (Gaussian) form.\footnote{Let us mention in this context
that the Gaussian dependence on the impact parameter $b$ for the
Drell-Yan process was already noticed by Collins and Soper \cite{CS81}
in their Sudakov analysis.}
However, and most importantly, with our method we can go beyond their
analysis and specify the absolute normalization of the power correction
that cannot be fixed within perturbation theory.
In our Drell-Yan calculation, this coefficient, see
Eq.~(\ref{eq:leadpow}), can be computed explicitly, and it turns out to
have the {\it opposite} sign relative to theirs and depend
logarithmically on the impact parameter.
The upshot is that the inclusion of power corrections leads to an
enhancement of the pion wave function in $b$ space, counteracting
partly this way the suppression provided by the familiar Sudakov
factor, similar in this respect to the observation made in
\cite{SSK00}, with the endpoint region $b\Lambda \sim 1$ (where
$b\Lambda$ is not a small expansion parameter and therefore
Eq.~(\ref{eq:DYpow}) becomes inaccurate) being less enhanced
relative to small $b$ transverse distances (cf. Fig.~\ref{fig:power2}).

%
\bigskip

\noindent
{\bf 4. Conclusions} \\
\indent We have focused on two specific cases -- the pion form
factor at leading power in $Q^{2}$ and the Drell-Yan process --
which expose all the salient features of the proposed
analytization methodology. On account of analytization of hadronic
functions as a whole, the dispersive conjugate of the running
coupling is defined unambiguously. Moreover, and even more
important, one can calculate not only the power of power
corrections to hadronic processes, but also their concomitant
coefficients because this approach does not contain an IR
renormalon ambiguity from the outset. In this way, we were able to
compute explicitly the first power correction in
$\frac{Q^{2}}{\Lambda ^{2}}$ to the pion form factor, as well as a
Sudakov-type factor to the Drell-Yan cross-section which contains
the leading power correction in $b^{2}\Lambda ^{2}$. Further
applications and phenomenological implications of our approach
will be pursued in forthcoming publications.
\bigskip

\acknowledgments
We wish to thank Alexander Bakulev, Johannes Bl\"umlein, Christos
Ktorides, Sergey Mikhailov, Dmitri V. Shirkov, and Wolfram Schroers
for stimulating discussions.
This work was supported in part by travel grants by the COSY
Forschungsprojekt J\"ulich/Goeke (A.I.K.) and Athens University
(N.G.S.).
\newpage   

\end{document}